\documentclass[a4paper]{article}
\pdfoutput=1
\usepackage{INTERSPEECH2020}
\usepackage{amsmath,amssymb}
\usepackage{multirow,booktabs}
\usepackage{graphicx}

\title{Deep F-measure Maximization for End-to-End Speech Understanding}
\name{Leda Sar{\i}, Mark Hasegawa-Johnson}
\address{
  Department of Electrical and Computer Engineering,\\
  University of Illinois at Urbana-Champaign, IL, USA}
\email{\{lsari2,jhasegaw\}@illinois.edu}

\begin{document}

\maketitle
\begin{abstract}
Spoken language understanding (SLU) datasets, like many other machine learning datasets, usually suffer from the label imbalance problem. Label imbalance usually causes the learned model to replicate similar biases at the output which raises the issue of unfairness to the minority classes in the dataset. In this work, we approach the fairness problem by maximizing the F-measure instead of accuracy in neural network model training. We propose a differentiable approximation to the F-measure and train the network with this objective using standard backpropagation. We perform experiments on two standard fairness datasets, Adult, and Communities and Crime, and also on speech-to-intent detection on the ATIS dataset and speech-to-image concept classification on the Speech-COCO dataset. In  all four of these tasks, F-measure maximization results in improved micro-F1 scores, with absolute improvements of up to 8\% absolute, as compared to models trained with the cross-entropy loss function.  In the two multi-class SLU tasks, the proposed approach significantly improves class coverage, i.e., the number of classes with positive recall. 
\end{abstract}

\section{Introduction}
Many machine learning datasets have a label imbalance or dataset bias problem. In many cases, either data is harder to collect for certain classes or the data collection phase is biased itself such that bias is introduced to the collected dataset. Typical training algorithms, optimized in order to minimize error, tend to do so by exacerbating bias, e.g., by providing higher recall and precision to the majority class than to minority classes. Therefore, the label imbalance problem raises the concern about fairness of machine learning systems in general~\cite{angwin2016machine,chouldechova2017fair,hardt2016equality}. Spoken language understanding (SLU) problems often suffer from label imbalance, in ways that may hide important errors from the designers of SLU systems.

Consider an SLU dataset such as Air Traffic Information Systems (ATIS) \cite{hemphill1990atis} and the speech-to-intent detection problem on this dataset.  About 75\% of the dataset carries the intent of searching for a flight, while conversely, some minority intent classes are represented by only a single training example; this is a severe label imbalance problem. Suppose that we train a model without any concerns about fairness or imbalance. The model will very likely learn to output the `flight' intent all the time, which will give us an accuracy of 75\% which is not low and could be acceptable depending on the application. Considering that there are roughly 30 classes in the whole dataset, one class will have a recall of 1.0 and precision of 0.75 and the remaining 29 classes will have both recall and precision of 0.0. In such a scenario, the F-measure, which is a harmonic average of precision and recall, will be 0.86 for the most common class and 0.0 for the rest, which will give an average of 0.03 which is not acceptable in many cases. 

There has been recent interest in introducing fairness to training in the machine learning literature \cite{kearns2017preventing,cotter2018training,jiang2019identifying}. Most such studies are applied to benchmark datasets related to socioeconomic problems, e.g., disparate impact~\cite{feldman2015certifying} or equal opportunity \cite{hardt2016equality}. In most such studies, fairness is defined to be the task of protecting against the use of explicit or implicit information about a protected attribute (e.g., gender  or race) in the decisions of the machine learning algorithm, for instance, framing the problem as a constrained optimization problem by introducing several penalties~\cite{zafar2015fairness,goh2016satisfying}. In this work, we introduce fairness into a speech-related problem, namely SLU. We also propose a positive and generalized definition of fairness, in terms of the missed detection and false alarm error rates suffered by all classes, regardless of whether the class definitions are matters of socioeconomic importance or merely engineering convenience.

There have been several studies on F-measure maximization \cite{nan2012optimizing,busa2015online,jansche2005maximum,waegeman2014bayes,decubber2018deep,jasinska2016extreme}. These models usually focus on binary classification using non-neural-network models: a situation in which the problem of F-measure optimization reduces to the problem of learning a threshold on the scores computed by the model to make a decision. We are aware of one study~\cite{decubber2018deep} that performs F-measure optimization for convolutional neural networks, but again, using a system that generates several binary classification outputs in parallel; in this scenario, F-measure optimization reduces to the task of tuning the thresholds of individual binary classifiers in order to maximize a weighted log likelihood. However, true multi-class classification, using the softmax output of the neural network, requires a modified definition of the F-measure.  There is no threshold that can be tuned; instead, F-measure optimization requires optimizing the model itself to generate `better' scores in terms of the F-measure. Model versus threshold optimization is the fundamental difference between this study and the previous ones.

In this work, our goal is to design a loss function to maximize the F-measure instead of the accuracy for DNNs. Our methods are tested on two standard socioeconomic classification problems from the literature on fairness (The UCI~\cite{Dua:2019} Adult \cite{kohavi1996scaling} and Communities and Crime \cite{redmond2002data} tasks), and on two SLU tasks (intent classification in ATIS, and detection of the named object in spoken captions that name only one object from the Speech-COCO dataset \cite{havard17coco}).  On the SLU  tasks,  we perform end-to-end SLU, i.e., we directly map speech input to the labels instead of performing automatic speech recognition (ASR) followed by natural language processing (NLP).  We pose the SLU problems as multi-class classification tasks and use the softmax output from the DNN, making it possible to apply the same optimization criterion to both the socioeconomic and SLU learning problems. We approximate the F-measure with a differentiable function of the softmax activations so that we can use the standard backpropagation algorithm~\cite{rumelhart1985learning} to train the DNN.

\section{Deep F-measure Maximization}
In this section, we will review the $F_\beta$-measure and present our proposed method.
\subsection{The $F_\beta$ Measure}
First, consider the binary classification problem. Given the true positive ($TP$), false positive ($FP$) and false negative ($FN$) counts for a test dataset, precision ($\mathrm{Prec}$) and recall ($\mathrm{Rec}$)  of the model can be written as follows: 
\begin{align}
\mathrm{Prec} =\frac{TP}{TP + FP}  \quad \text{and} \quad \mathrm{Rec} = \frac{TP}{TP + FN}.
\end{align}
Given these definitions, $F_\beta$ measure is defined as a weighted harmonic mean of precision and recall~\cite{van1974foundation}  
\begin{equation}
	F_\beta = \frac{(1+\beta^2) \mathrm{Prec} \cdot  \mathrm{Rec}}{ \beta^2\mathrm{Prec}+ \mathrm{Rec}}.
\end{equation} 
If we substitute the precision and recall expressions to the above equation, we can also write the $F_\beta$ measure as 
\begin{equation}
F_\beta = \frac{(1+\beta^2)TP}{ \beta^2(TP + FN)+ (TP+FP)}.
\end{equation} 

For the multi-class classification case, there are several ways of computing the $F_\beta$-measure. We can compute the average precision and recall over all classes and then take their harmonic mean to get the micro-$F_\beta$-measure. Alternatively, we compute the class-wise $F_\beta$-measures and take the average over classes to get the average-$F_\beta$-measure. In this work, we optimize the latter. Suppose that there are $K$ classes and $N_k$ denotes the number of data points from class $k$, then the average $F_\beta$ is computed as
\begin{equation}
	F_\beta = \frac{1}{K} \sum_{k=1}^{K} \frac{(1+\beta^2)TP(k)}{ \beta^2N_k+ (TP(k)+FP(k))}. \label{eq:Fbeta}
\end{equation}
 Note that the $N_k$ term corresponds to ($TP(k) + FN(k)$). 
 
\subsection{Empirical Optimization of $F_\beta$}
Earlier works on $F_\beta$-measure have focused on learning a threshold for making a decision for the binary classification problem. On the other hand, in the case of multi-class classification with DNNs, the class decision is made by taking the softmax at the output layer and then by choosing the class with the highest softmax activation. Therefore, in $F_\beta$ maximization with neural networks we do not aim at identifying the threshold but designing a loss function that is differentiable so that we can use the backpropagation method to learn the DNN model parameters.

Eq.~\eqref{eq:Fbeta}
contains counting which is expressed using indicator functions that are not differentiable. For example, given that the softmax activations for the $n$\textsuperscript{th} data point, or token, are $q_n(k), ~ k=1,2,\cdots, K$ and that $p_n$ is the one-hot representation of the true label, the number of true positives for a certain class $k$ is written as 
\begin{equation}
	TP(k) = \sum_n  \mathbf{1}[\arg \max p_n =k \wedge \arg \max q_n =k ]
\end{equation}
where the indicator function $\mathbf{1}$ is not differentiable. Therefore, we need a differentiable approximation for $F_\beta$. To achieve this, instead of the hard counts, we use the soft counts which are obtained from the softmax activations. To make the largest activations equal to 1, we do the following normalization on the activations for each token:
\begin{equation}
	q_n' = \frac{q_n}{\max_k q_n(k)}.
\end{equation}
Using these soft counts, we approximate the terms in Eq.~\eqref{eq:Fbeta} as 
\begin{align}
	TP(k) &\approx \sum_{n \in S_k}  q_n'(k) \label{eq:tp}\\
	TP(k) + FP(k) &\approx \sum_{n \in S} q_n'(k) \label{eq:tpfp}
\end{align}
where $S_k$ denotes the set of indices for data tokens with label $k$ and $S$ is the set of all indices in the dataset. We do not approximate $N_k$ as it is determined directly from the dataset. Thus, our loss function becomes the negative of the approximate $F_\beta$:
\begin{equation}
	\mathcal{L} = - \frac{1}{K} \sum_{k=1}^{K} \frac{(1+\beta^2)\sum_{n \in S_k} 	q_n'(k)}{ \beta^2N_k+ \sum_{n \in S} 	q_n'(k)}. \label{eq:loss}
\end{equation}
Since $q_n'$ is a differentiable function of $q_n$, it is also differentiable with respect to the DNN model parameters. Hence, we can learn the network weights by backpropagating the derivatives of the loss function in Eq.~(\ref{eq:loss}).  The loss function in Eq.~(\ref{eq:loss}) is not specific to fully-connected neural networks but can be used for any neural network with a softmax output layer. 

In the approximations given in Eqs.~\eqref{eq:tp} and~\eqref{eq:tpfp}, instead of $q'$, we could have used $q$ directly, or we could have computed the softmax by first scaling the pre-softmax activations by a constant to increase the sharpness of the final activations. However, in our experiments, we saw that the approximations proposed in the equations above performed the best. 

\section{Experiments}
In this section, we will describe two sets of experiments. Although our main focus will be on dealing with dataset bias in SLU systems, the first set of experiments will be on smaller datasets for non-speech, binary classification tasks. These are usually used as benchmark tasks as they reflect some societal bias. The second set of experiments will be on speech-to-intent and speech-to-concept classification which are both multi-class classification tasks. Details of the models and the results will be presented in the following subsections. 

\subsection{Experiments on Socioeconomic Data}
The first set of experiments are performed on non-speech tasks. The goal here is to show whether the proposed method is providing any gains as compared to cross-entropy based training. Since the dataset bias is usually discussed in the realm of socioeconomic data with certain protected attributes such as race, gender, age-group etc., we first want to investigate whether we achieve an improvement in these tasks. 
\begin{table}
	\centering
	\caption{Binary classification performance on two UCI datasets}
	\label{table:uci}
	\resizebox{\columnwidth}{!}{%
	\begin{tabular}{llccccc}
		\toprule
		Data& Loss & Prec & Rec & Micro-$F_1$ & Avg-$F_1$ &  Accu.\\ \midrule
		\multirow{2}{*}{Adult} & xent & 0.7977& 0.6193& 0.6973 & 0.6389 & 0.8085 \\
		& deepF & 0.8196  & 0.6170 & 0.7040 & 0.6361 & 0.8107 \\ 
		\midrule
		\multirow{2}{*}{C\&C} & xent  & 0.7422	& 0.7075 & 0.7245 & 0.7206 & 0.7940 \\
		& deepF & 0.7541	& 0.7319 & 0.7428	& 0.7413	& 0.8040 \\
		\bottomrule
	\end{tabular}
	}%
\end{table}
For this task, we use two datasets from the UCI repository~\cite{Dua:2019}, namely, Adult \cite{kohavi1996scaling} and Communities and Crime \cite{redmond2002data}.
In the Adult dataset, given the personal attributes (age, race, marital status, education level, etc.) of a person, the goal is to estimate whether the person has an income over \$50K/year. The majority class, i.e. individuals with income less than \$50K/year, comprises 76\% of the data points. 
In the Communities and Crime (C\&C) dataset, the goal is to detect if a community has a high crime rate where, as described in \cite{kearns2017preventing,cotter2019training}, we define `high crime rate'  to mean a crime rate above the $70^{\textrm{th}}$ percentile of the training dataset.
The majority class, i.e., low crime-rate, comprises 70\% of the samples.

Both the Adult and C\&C tasks are two-class problems, for which a standard F-measure is well-defined. Our interest is the maximization of a multi-class F-measure, therefore the F-measures of both majority and minority classes are first computed, and then averaged as shown in Eq.~(\ref{eq:loss}).

In both tasks, we use fully-connected neural networks with 16 units per layer. The number  of layers are 7 and 4 for the Adult, and C\&C datasets, respectively. The output is a softmax layer with 2 units. As a baseline, we use the models trained with cross-entropy loss and compare them to models trained by the proposed deep $F_\beta$ loss. Table~\ref{table:uci} shows the average precision, average recall, micro-$F_1$ and classification accuracy for both cross-entropy model (xent) and the proposed model (deepF) for both datasets where we take $\beta=1$. For both datasets, we improve the micro-$F_1$ and accuracy. For the C\&C dataset, we also see improvement in the average-$F_1$ score. 

\subsection{Experiments on Spoken Language Understanding}
The second set of experiments are on speech related tasks. We investigate direct speech-to-meaning systems where instead of the conventional two-step process (ASR+NLP), our goal is to directly understand the speech signal in an end-to-end framework.  
For the SLU problem, we run experiments on two tasks: speech-to-intent detection, and speech-to-concept classification; both of which are multi-class classification problems. We work on the ATIS dataset~\cite{hemphill1990atis} for the speech-to-intent task, where the intents are `searching for a flight', `getting airport information', `local transportation options', etc. There are 29 intents in the whole dataset 8 of which do not appear in the training set.
For the speech-to-concept task, we use the Speech-COCO dataset~\cite{havard17coco}. This dataset consists of synthesized speech signals for the image captions in the MS-COCO dataset~\cite{lin2014coco}. We define the task to be mapping the spoken image captions to the image label. There are 80 classes in the dataset. 

\begin{table}
	\centering
	\caption{Number of classes and the frequency (in \%) of the most frequent top-3 classes for ATIS and Speech-COCO datasets based on the training data}
	\label{table:label-freq}
	\begin{tabular}{lcccc} 
		\toprule
		Data & \#Classes & Top1  & Top2& Top3 \\
		\midrule
		ATIS & 29 & 73.7& 8.5& 5.1 \\ 
		Speech-COCO & 80 & 22.6 & 3.5 & 3.1\\ 
		\bottomrule
	\end{tabular}
	\vspace{-10pt}
\end{table}
In Table~\ref{table:label-freq}, we show the number of classes and the frequency of the most common three labels in both ATIS and Speech-COCO training sets. As shown in this table, the classes are highly imbalanced and we have dataset bias. Given these statistics, a model that always predicts the majority class will have 73.7\% and 22.6\% accuracy on the ATIS and Speech-COCO training datasets, respectively. If we compute the micro-F1 for such models, they will be 0.0293 for ATIS and 0.0046 for Speech-COCO which are very low (less than $3\%$) and these numbers will get even lower for datasets with more classes. Especially, in the ATIS case, we see that relatively high accuracy does not necessarily mean a classifier that is fair to all classes. 

\begin{figure}
	\centering
	\includegraphics[width=0.7\columnwidth]{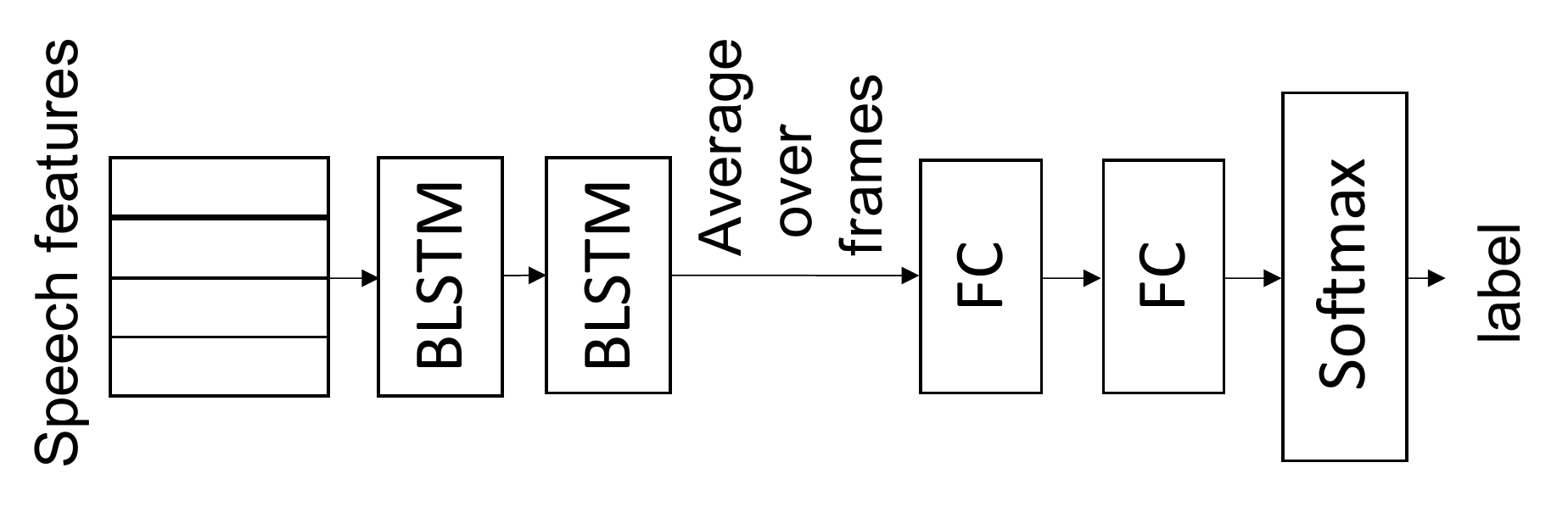}  
	\caption{Our end-to-end SLU architecture based on \cite{sari2020training}}
	\label{fig:e2eSLU}
	\vspace{-10pt}
\end{figure}

End-to-end SLU has gained interest as a means to overcome the error propagation problem, in which speech transcription errors cause speech understanding errors \cite{qian2017exploring,serdyuk2018towards,haghani2018audio,caubriere2019curriculum,lugosch2019speech,sari2020training}. This work uses the speech branch of the multiview model described in~\cite{sari2020training} which consists of a BLSTM based encoder and a classifier with fully-connected layers (Fig.~\ref{fig:e2eSLU}).
Since our focus is on designing the loss function for F-measure maximization, we keep the DNN architecture otherwise identical to that used in~\cite{sari2020training},
and use speech-only training instead of the multi-task training protocol described in~\cite{sari2020training}.
For ATIS experiments, the model has a single BLSTM layer with 128 units and two fully-connected layers with 64 units each.  
For Speech-COCO experiments, the model has 2 BLSTM layers with 128 units each and two fully-connected hidden layers with 128 and 64 nodes. The dataset comes with train and validation splits; we reserve 25\% of the training subset as our development set.  In both cases, we experiment with ReLU and leaky ReLU non-linearity for the fully-connected layers, we set the learning rate to 0.001, and we use Adam optimizer. 

\begin{table*}
	\centering
	\caption{Multi-class classification performance (precision, recall, micro-F1, average-F1, accuracy and coverage) on end-to-end SLU problems for different models (M1: ReLU nonlinearity, M2: leaky ReLU nonlinearity)}
	\label{table:atis+coco}
	\resizebox{2\columnwidth}{!}{%
	\begin{tabular}{llcccccc|cccccc}
	 	\toprule
	 	 & \multicolumn{7}{c|}{M1 - ReLU nonlinearity} & \multicolumn{6}{c}{M2 - leaky ReLU nonlinearity} \\
	 	Data & Loss & Prec & Rec & Mic-$F_1$ & Avg-$F_1$ &  Accu & C & Prec & Rec & Mic-$F_1$ & Avg-$F_1$ &  Accu & C\\ 
	 	\midrule 
	 	\multirow{2}{*}{ATIS} & xent  & 0.0244 & 0.0345 & 0.0286 & 0.0286 & 0.7772 & 1 & 0.0313	& 0.0362 &	0.0336	& 0.0332 &0.6697 & 2 \\
	 & deepF & 0.0520 & 0.0554 & 0.0536 & 0.0516 & 0.6484 & 4 & 0.1054 & 0.0936 & 0.0991& 0.0947 & 0.7447 & 5 \\
	 	\midrule 
	 	\multirow{2}{*}{COCO} & xent  & 0.1992 & 0.2268	& 0.2121 & 0.1956 & 0.3538& 50 & 0.3876 & 0.3716 & 0.3794 & 0.3509 & 0.4473 & 74 \\
	 	& deepF  & 0.2539 & 0.3137 & 0.2807 & 0.2676 & 0.3264 & 79 &  0.3927 & 0.3994 & 0.3960 & 0.3895 & 0.4439 & 79\\
	 	\bottomrule  
	\end{tabular}
	}
\end{table*}

In Table~\ref{table:atis+coco}, we show the average precision, average recall, micro-$F_1$, average-$F_1$, accuracy and coverage. We define the coverage as the number of classes with non-zero recall. This is an indicator of fairness as it highlights the very low number of classes that have non-zero recall under a standard cross-entropy training paradigm. We report the results on both ATIS and Speech-COCO datasets.  Training with cross-entropy loss is compared to training with the proposed $F_\beta$ measure (with $\beta=1$). We first experiment with model 1 (M1) that has ReLU non-linearity. For both datasets, we see that deep F-measure maximization (deepF) results in higher micro-$F_1$ and average-$F_1$ as compared to the cross-entropy (xent) model. In both cases, we also see that we increase the coverage significantly. Especially, on the ATIS dataset, we see that the cross-entropy model only outputs the majority class label. On the other hand, the deepF model has a coverage of 4 which shows that it is able to output labels from different classes. On the Speech-COCO dataset, with the deepF model, we cover almost all classes (79 out of 80). However, we also observe that there is a trade-off between coverage and accuracy. While trying to cover different classes, the model misses some of the majority class data points which leads to slightly lower accuracy as compared to the cross-entropy model. This is an expected outcome as 
the deep F-measure optimization aims at achieving better F-measure without paying attention to the overall accuracy. If our goal is fairness, and if the difference in accuracy is not large, deepF may still be the preferred approach. When we trained M1 for larger $\beta$ (more emphasis on recall), we saw that ReLU neurons start to die and hence lead to the degenerate solution, i.e., outputting the majority class label. Therefore, we also perform experiments with leaky ReLU (model 2, M2). With M2, we observe better baselines with the cross-entropy objective. However, our previous conclusions still hold, deepF leads to higher F-measure and increased coverage. 

\begin{figure}[t!]
	\centering
	\includegraphics[width=0.9\columnwidth]{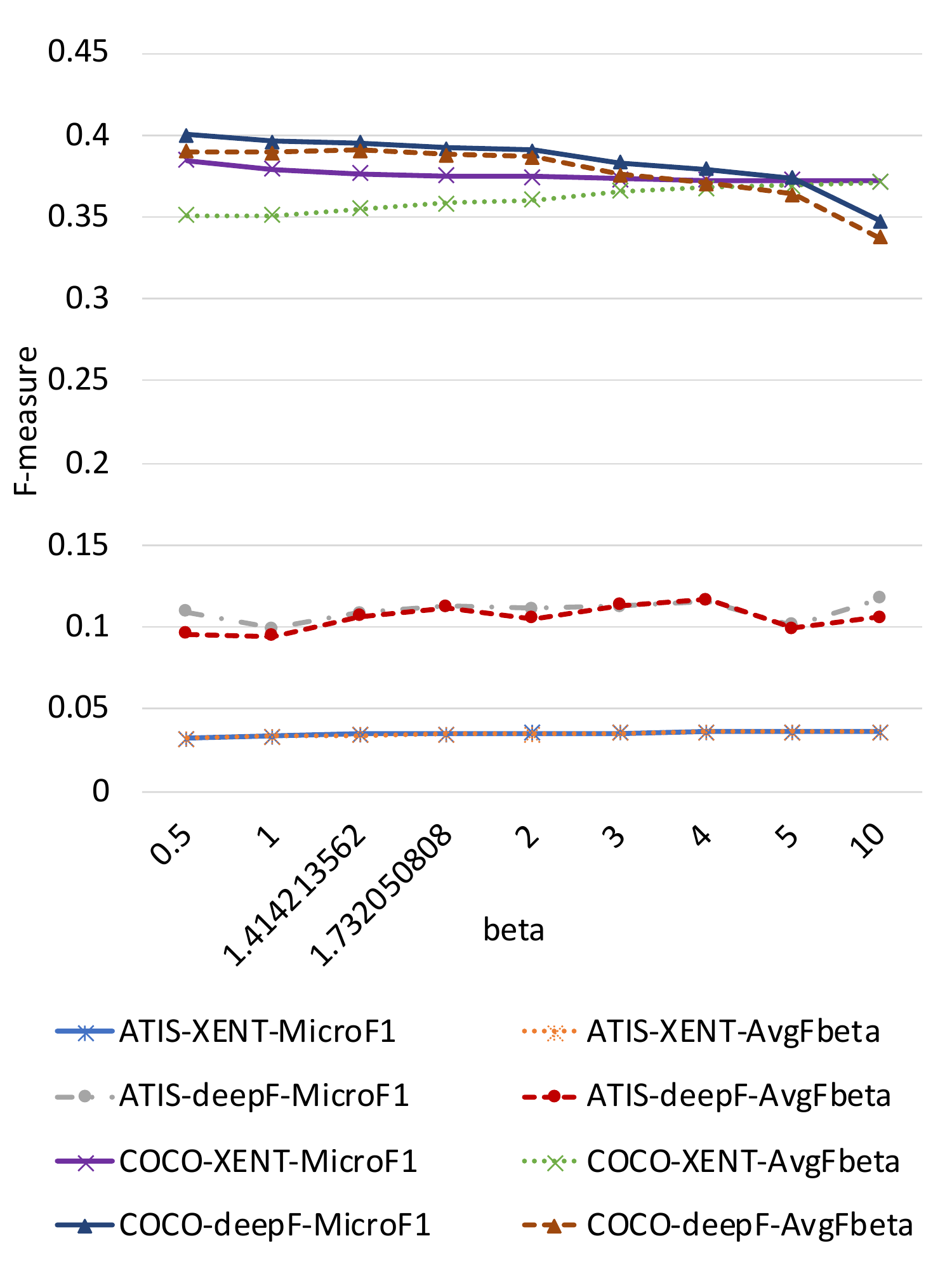}
	\caption{Micro-$F_1$ and average-$F_\beta$ values on the ATIS and Speech-COCO datasets for different $\beta$ after training with cross-entropy (XENT) or deep F-measure (deepF) losses}
	\label{fig:atis-beta}
	\vspace{-10pt}
\end{figure}

In Fig.~\ref{fig:atis-beta}, we show the average-$F_\beta$ and micro-$F_1$ obtained from M2 for ATIS and Speech-COCO datasets, for different values of $\beta$. Note that in the case of cross-entropy training, we only train a single model, then compute its $F_\beta$ for different values of $\beta$. On the other hand, we train a model
for each $\beta$ in the case of deep F-measure maximization. The cross-entropy system is trained for 25 epochs. The deep-F system is trained for 15 epochs using cross-entropy, then for 10 epochs using the $F_\beta$ measure.

Results on the ATIS dataset (lower half of the results in Fig.~\ref{fig:atis-beta}) show that the proposed deep F-measure maximization approach leads to 6-8\% absolutely higher micro-$F_1$ and average-$F_\beta$ as compared to the cross-entropy model for a wide range of $\beta$.  By comparing M2 results in Table~\ref{table:atis+coco} to Fig.~\ref{fig:atis-beta}, it is possible to compare the sizes of the improvements in coverage (about 3-fold improvement at $\beta=1$) and in $F_1$.  Micro-$F_1$ improves by a factor of 2.9 at $\beta=1$, and by a factor of 3.2 at $\beta=4$ (from 0.0359 to 0.1161). These results suggest that increasing coverage 
has a large (up to 8\% absolute) effect on the micro-$F_1$.  

As shown in upper half of the Fig.~\ref{fig:atis-beta}, for the Speech-COCO dataset, F-measures are around 35-40\%. On this dataset, deep F-measure maximization still performs better (up to 5\% absolute) than the cross-entropy loss when $\beta<4$ and there is not a significant difference in the F-measure for different $\beta$. However, when $\beta \ge 4$, the performance starts to fall below the cross-entropy model. Still, if we look at the coverage for these models, we see that it is 79 which is higher than that of the cross-entropy model. This means that we have nonzero recall for more classes but the individual F-measures per class are, on average, lower than their cross-entropy counterparts.

\section{Conclusions and Future Work}
In this work, we proposed a method to maximize the F-measure while training a DNN to deal with the label imbalance problem that is frequently encountered in many datasets. We approximated the average $F_\beta$ using soft counts obtained from the softmax activations of the DNN. We compared our proposed method to cross-entropy based training in our experiments. We showed that this method can be applied to different types of DNNs, either fully-connected or BLSTM based, as long as their final layer is a softmax layer. In our experiments on two SLU problems, namely the ATIS speech-to-intent detection problem and the Speech-COCO speech-to-image label classification task, we showed that deep F-measure maximization performs better than the cross-entropy model in terms of micro-$F_\beta$, average-$F_\beta$ and the coverage of classes. Especially, significantly increased coverage shows that the proposed method provides a fair way of treating minority classes.

There are several future directions for research. One direction is to deal with the coverage versus accuracy trade-off, e.g., to explore multi-task or constrained learning methods that might improve coverage and fairness without harming performance for the majority class. Another issue that we would like to address is the performance degradation for high $\beta$ cases for Speech-COCO. 
We also would like to perform experiments on larger datasets with real speech instead of synthesized speech. 

\section{Acknowledgments}
The authors would like to thank Samuel Thomas from IBM Research for helping with preparing the ATIS dataset. The authors would also like to thank the IBM-ILLINOIS Center for Cognitive Computing Systems Research (C3SR) - a research collaboration as part of the IBM AI Horizons Network. The authors are partially supported by the National Science Foundation under Grant No. NSF IIS 19-10319. Any opinions, findings, and conclusions or recommendations expressed in this material are those of the authors and do not necessarily reflect the views of the National Science Foundation.

\newpage
\bibliographystyle{IEEEtran}
\bibliography{main}

\end{document}